\documentclass[aps,pra,twocolumn,groupedaddress]{revtex4-1}
\usepackage{color}
\usepackage{dcolumn}
\usepackage{graphicx}
\usepackage{amsmath}
\usepackage{amssymb}

\begin{document}


\title{Probing photon correlations in the dark sites of geometrically frustrated cavity lattices \\}


\author{W. Casteels, R. Rota, F. Storme, C. Ciuti}
\affiliation{Universit\'{e} Paris Diderot, Sorbonne Paris Cit\'{e}, Laboratoire Mat\'{e}riaux et Ph\'{e}nom\`{e}nes Quantiques,
CNRS-UMR7162, 75013 Paris, France}


\date{\today}

\begin{abstract}

We theoretically explore the driven-dissipative physics of geometrically frustrated lattices of cavity resonators with relatively weak nonlinearities, i.e. a photon-photon interaction smaller than the loss rate.
In such systems, photon modes with zero probability at  'dark' sites are present at the single-particle level due to interference effects.  In particular, we study the behavior of a cell with three coupled resonators as well as extended Lieb lattices in 1D and 2D. By considering a partial pumping scheme, with the driving field  not applied to the dark sites, we predict
that even in presence of relatively weak photon-photon interactions the nominally dark sites achieve a finite photonic population with strong correlations.
We show that this is a consequence of biphoton and multiphoton states that in the absence of frustration would not be visible in the observables. 
\end{abstract}

\pacs{}

\maketitle

\section{Introduction}
In recent years driven-dissipative nonlinear photonic systems have gained a lot of interest to study the quantum many-body physics of light (see for example Refs. \cite{RevModPhys.85.299, ANDP:ANDP201200261, RevModPhys.86.1391} for recent reviews). Experimentally these systems can be realized, e.g., 
in semiconductor optical microcavities or superconducting quantum circuits at microwave frequencies. Since these systems are subject to losses, a continuous drive has to be applied in order to reach a non trivial steady-state. This configuration provides various important opportunities such as the fundamental study of systems out of equilibrium. Moreover, it is in principle possible to engineer the coupling with the environment, either through the losses or the drive, in order to accompany the system to a steady-state of interest \cite{Diehl:2008aa}. 

An important class of lattices of nonlinear cavities is described by the driven-dissipative Bose-Hubbard model. Experimentally it can be implemented with arrays of zero-dimensional microcavities with a Kerr nonlinearity, coupled by photon tunneling. Recently there have been various proposals for the realization of strongly correlated states with these systems \cite{PhysRevLett.103.033601,Gerace:2009aa, PhysRevLett.104.183601, PhysRevLett.104.113601,PhysRevA.83.021802,1367-2630-15-2-025015,PhysRevLett.108.206809, PhysRevLett.108.233603,PhysRevLett.110.233601,PhysRevA.87.053846, doi:10.1142/S0217979214410021,PhysRevA.91.053815, PhysRevA.90.043822}. A class of emerging systems are those with geometric frustration: these lattices can have single-particle eigenstates with dark sites, i.e. zero photon occupation. The frustration can induce a flat band with vanishing kinetic energy in the single-particle energy spectrum. In general it is well known that frustration and a flat energy band can lead to correlated states, for example in the context of magnetism \cite{doi:10.1146/annurev.ms.24.080194.002321,Morris16102009,RevModPhys.85.1473}, fractional Chern insulators \cite{PhysRevLett.106.236804,doi:10.1142/S021797921330017X, Parameswaran2013816} and ultra-cold atoms \cite{PhysRevLett.99.070401, PhysRevA.82.041402}. In particular the Lieb lattice (see Fig. \ref{Fig1}) has recently received a lot of interest and various experimental realizations have been presented with photonic lattices \cite{1367-2630-16-6-063061, PhysRevLett.114.245503, PhysRevLett.114.245504, 2015arXiv150505652B}. In the presence of a strong nonlinearity the flat band can give rise to an incompressible polariton gas \cite{PhysRevLett.115.143601}. For the 2D Lieb lattice a rich topological structure has been predicted in the presence of a gauge field \cite{PhysRevB.54.R17296,PhysRevB.82.085310, PhysRevA.83.063601, PhysRevB.87.125428}.

In this paper we show that a partial driving scheme on geometrically frustrated lattices can give rise to strong correlations at the nominally dark sites for relatively weak photon nonlinearities, within the reach of experimental platforms based, e.g., on semiconductor optical microcavities. First we consider a system consisting of a cell of three coupled resonators and describe analytically and numerically the key paradigm. Then we explore the extended Lieb lattice in 1D and 2D by using the corner-space renormalization method \cite{PhysRevLett.115.080604}. We demonstrate that in presence of geometric frustration, the physics at the nominal dark sites is dominated by multiphoton resonances even in presence of moderate nonlinearity. 

The article is organized as follows: in Section II we present the theoretical model and the general problem. In Section III we discuss our results for the case of three coupled cavities (Subsection A) and then for extended 1D and 2D Lieb lattices (Subsection B). Finally, Section IV is devoted to conclusions and perspectives. 

\section{Theoretical framework}

\begin{figure}[h]
\begin{center}
  \includegraphics[scale=0.75]{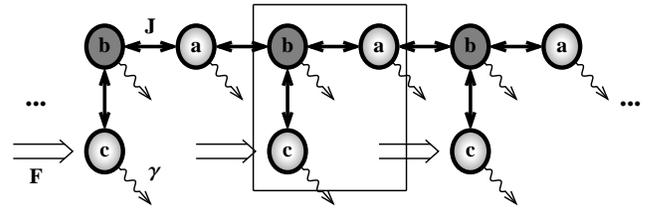}
  \end{center}
  \caption{\label{Fig1} Sketch of the driven-dissipative one dimensional Lieb lattice with the partial driving scheme. The elementary unit cell is indicated by the box. Only the $c$-sites are coherently driven and the $b$-sites are dark with respect to the $a$- and $c$-sites. The indicated system parameters are discussed in the text.}
\end{figure}

Interacting bosons on the one-dimensional Lieb lattice with a partial drive (as presented schematically in Fig. \ref{Fig1}) can be described with the following tight-binding Hamiltonian ($\hbar = 1$ in the following):
\begin{eqnarray}
\hat{H} &&=  -J\sum_{i}\left(\hat{a}_i^\dagger\hat{b}_i+\hat{b}_i^\dagger\hat{c}_i+ \hat{a}_i^\dagger\hat{b}_{i+1} + h.c.\right) \nonumber \\
&&+  \sum_{i,s \in \{a,b,c\}}\left(\omega_c\hat{s}_i^\dagger\hat{s}_i+\frac{U}{2}\hat{s}_i^\dagger\hat{s}_i^\dagger\hat{s}_i\hat{s}_i \right) \nonumber \\
&&+ \sum_{i}\left( Fe^{-i\omega_pt}\hat{c}_i^\dagger+F^*e^{i\omega_pt}\hat{c}_i \right),   
\end{eqnarray}  
The operators $\{\hat{a}_i,\hat{b}_i,\hat{c}_i\}$ are the annihilation operators for photons on the different sites. The integer index $i$ is used to denote the unit cells and $s \in \{a,b,c\}$ the different sites, as indicated in Fig. \ref{Fig1}. The first line represents the hopping between the different sites, $J$ being the corresponding coupling. The second line describes the on-site part with the resonator frequency $\omega_c$ and a photon-photon interaction strength $U$. Finally, the last line represents a coherent drive on the $c$-sites (see sketch in Fig. \ref{Fig1}) with amplitude $F$ and frequency $\omega_p$. 

For the non-interacting case ($U=0$) the Hamiltonian of the closed system (without the drive, i.e. $F = 0$) can be diagonalised in the reciprocal space, resulting in three energy bands with a flat band in the middle \cite{PhysRevA.87.023614, PhysRevLett.115.143601}. The eigenstates corresponding to the flat band contain no occupation on the $b$-sites, which are dubbed dark sites. This is presented schematically in Fig. \ref{Fig1}.

The driven-dissipative dynamics is described by the Lindblad master equation for the time evolution of the density matrix $\hat{\rho}(t)$, given by:
 \begin{eqnarray}
\frac{\partial\hat{\rho}}{\partial t}= &i\left[\hat{\rho},\hat{H}\right] + \frac{\gamma}{2}\sum_{i,s}\left(2\hat{s}_i\hat{\rho}\hat{s}^\dagger_i - \hat{s}_i^\dagger\hat{s}_i\hat{\rho}-\hat{\rho}\hat{s}_i^\dagger\hat{s}_i \right),
\label{eq:Master}
\end{eqnarray}
where $\gamma$ is the cavity loss-rate. The steady-state properties of the system are determined by solving the master equation for $\frac{\partial\hat{\rho}}{\partial t}=0$. For our analysis the frame rotating at the drive frequency is considered, the Hamiltonian then becomes time-independent and the relevant parameter is the detuning between the pump and the cavity frequencies, namely $\Delta = \omega_p - \omega_c$. 

\begin{figure}[h]
\begin{center}
  \includegraphics[width=0.45 \textwidth]{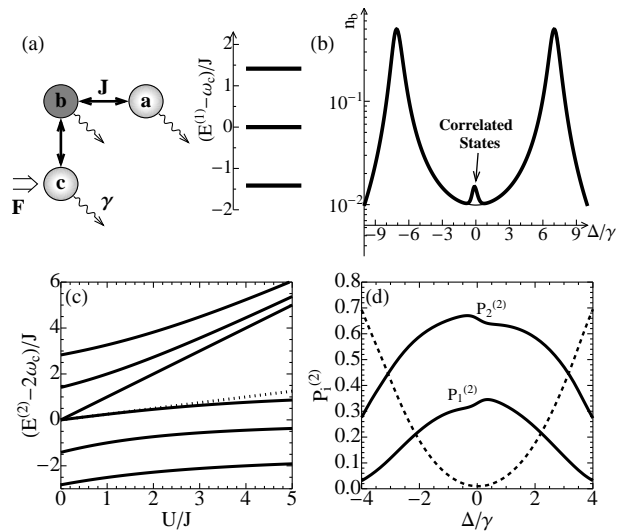}
  \end{center}
  \caption{\label{Fig10} (a) Sketch of three coupled driven-dissipative cavities (corresponding to a single unit cell of the Lieb lattice with open boundary conditions) and the corresponding single-particle energy level structure for the closed system (in units of $J$), with three levels at $E^{(1)} - \omega_c = -\sqrt{2}J$, $0$ and $\sqrt{2}J$. (b) The population on the central site  $n_b$ (on a log scale) as a function of the detuning $\Delta$ (in units of $\gamma$) for $F=\gamma$, $U=0.1\gamma$ and $J=5\gamma$. An anomalous peak is found around $\Delta=0$, completely missed by the non-equilibrium Gross-Pitaevkii approximation (thin line). (c) The energies of the six two-photon eigenstates (in units of $J$) as a function of the non-linearity $U$ (in units of $J$). The dotted line is the approximation $U/4$ for small $U/J$ discussed in the text. (d) The probabilities (normalized to the two-photon manifold) of the two-photon eigenstates (\ref{EqES}) in the limit $F \to 0$ as a function of the detuning. The dashed line gives the sum of the probabilities for the other four eigenstates. The system parameters are $J = 5\gamma$ and $U = 0.1\gamma$. }
\end{figure}

We would like to mention a very recent and impressive experimental realization of the photonic Lieb lattice with coupled micro-pillar optical cavities \cite{2015arXiv150505652B}, which has been studied in the case of non-resonant, incoherent excitation. The same system can be excited by quasi-resonant coherent drive, which is the configuration considered in this article.
Moreover, the second order correlation function is an experimentally accessible quantity with these systems  (see for example Ref. \cite{:/content/aip/journal/apl/107/22/10.1063/1.4936889}).

\section{Results}
\subsection{Three coupled cavities \label{Sec3Cav}}

First, we consider the simplest case of a single unit cell with open boundary conditions, corresponding to 3 coupled cavities with a coherent drive on one of the outer sites (see sketch in Fig. \ref{Fig10} (a)). To obtain the results, we have performed an exact numerical integration of the master equation. The results are compared with the non-equilibrium Gross-Pitaevskii approximation, obtained by replacing the operators by $\mathbb{C}$-numbers corresponding to neglecting all correlations, which is typically a good approximation for small nonlinearities and large densities \cite{RevModPhys.85.299}. The single-particle spectrum is also presented in Fig. \ref{Fig10} (a) and contains three states, with the middle one having no occupation of the central site. When driving the system on resonance with this eigenstate ($\omega_p = \omega_c$, i.e. $\Delta = 0$) and due to the partial drive scheme, a large coherent population of this eigenstate is expected, resulting in a relatively low density on the central site. This is shown in Fig. \ref{Fig10} (b) where the density on the central site $n_b = \langle \hat{b}^\dagger\hat{b}\rangle$ is presented as a function of the detuning for a relatively weak nonlinearity ($U=0.1\gamma$). For comparison the semiclassical Gross-Pitaevskii result is also presented in Fig. \ref{Fig10} (b) which gives a good approximation for the two resonances at $\Delta = \pm \sqrt{2}J$ . However, the central site density clearly exhibits a deviation around $\Delta \simeq 0$, where the exact result reveals an additional peak which is not at all captured by the Gross-Pitaevskii treatment. For the linear spectrum ($U = 0$) the population of the central site $n_b$ is due only to the population of the broadened resonances at $\Delta = \pm\sqrt{2}J$, resulting in a minimum at $\Delta = 0$, similar to the Gross-Pitaevskii result in \ref{Fig10} (b). This deviation at $\Delta \simeq 0$ is a clear indication that a relatively large part of the density at the central site is due to correlated states, neglected in the Gross-Pitaevskii treatment.

In order to gain insight into these correlated states, we examine the $6$ two-particle eigenstates of the closed system. Their energies are plotted in Fig. \ref{Fig10} (c) as a function of $U/J$. In the considered regime with a moderate nonlinearity two of these states are close to resonance for a detuning $\Delta = 0$, namely:    
\begin{subequations}
\begin{eqnarray}
& \vert \Psi^{(2)}_1 \rangle = \frac{1}{\sqrt{3}}\left(\left| 2,0,0 \right> - \left| 0,2,0 \right> + \left| 0,0,2 \right> \right),  \label{State1}\\
& \vert\Psi^{(2)}_2\rangle = \frac{1}{2\sqrt{6}}\left(| 2,0,0 \right> + \left| 0,0,2 \right> + 2\left| 0,2,0 \right> \nonumber \\
& - 3\sqrt{2}\left| 1,0,1 \right> ),
\label{State2}
\end{eqnarray}
\label{EqES}
\end{subequations}
where we used the Fock basis representation: $\left| n_{a}, n_{b}, n_{c}\right>$, with $n_{d}$ the number of photons on site $d$. Indeed, among the six two-photon states, these two are the most probable (see Fig. \ref{Fig10} (d) and discussed later in more detail). The first state in Eq. (\ref{State1}) is an eigenstate of the system for all parameters, with eigenenergy $E^{(2)}_1 = U$, while the second state in Eq. (\ref{State2}) is an approximation, with energy $E^{(2)}_2 = U/4$, that becomes exact in the limit $U/J \to 0$ (see Fig. \ref{Fig10} (c)). It is clear that populating these two-photon states would result in a contribution to the density at the central site.   

\begin{figure}[h!]
\begin{center}
  \includegraphics[width=0.5 \textwidth]{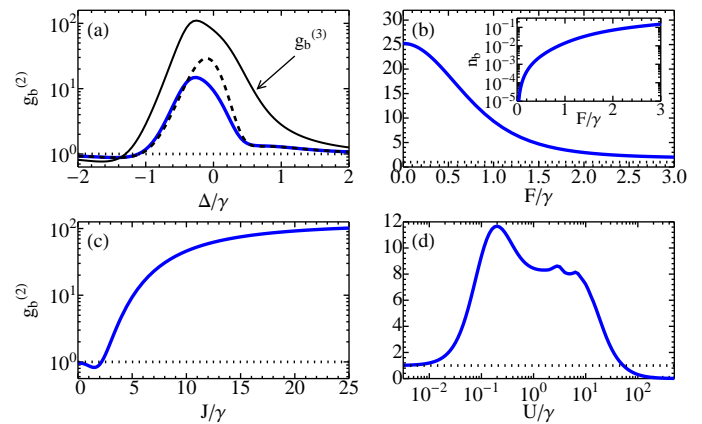}
  \end{center}
  \caption{\label{Fig5}The second-order correlation function $g_b^{(2)}$ on the central dark site as a function of the system parameters (in units of the loss rate $\gamma$), namely the detuning $\Delta$ (a), the driving amplitude $F$ (b), the hopping term $J$ (c) and the on-site photon-photon interaction $U$ (d) (note the logarithmic scales for the vertical axes in (a) and (c) and the horizontal axis in (d)). While not varied, the fixed parameters are $\Delta=0$, $F=\gamma$, $U=0.1\gamma$ and $J=5\gamma$. The dotted lines indicate the value $g_b^{(2)} = 1$ for a coherent state. In (a) the dashed line is the analytical result in the limit $F \to 0$ and the full black line is the result for $g_b^{(3)}$ which is also strongly peaked around $\Delta\simeq 0$. In (b) the inset presents the corresponding density on the dark site $n_b$.}
\end{figure}

The states presented in Eqs. (\ref{EqES}) both contain a probability for a double occupation of the dark site, so their presence is expected to influence the normalised second order correlation function on the dark site, defined as
$
g_b^{(2)} = \langle\hat{b}^\dagger\hat{b}^\dagger\hat{b}\hat{b}\rangle/\langle\hat{b}^\dagger\hat{b}\rangle^2.
$
This is indeed what is reported in Fig. \ref{Fig5} where $g_b^{(2)}$ is presented as a function of the system parameters. A super-bunching is revealed with a $g_b^{(2)}$ much larger than $1$, clearly demonstrating a break-down of the Gross-Pitaevskii approximation, which instead predicts $g^{(2)} = 1$ for all sites. Fig. \ref{Fig5} (a) presents $g_b^{(2)}$ as a function of the detuning showing a large peak around zero detuning. This is a consequence of the coherent single-particle population being at a minimum here while the correlated states (\ref{EqES}) are close to resonance. From Fig. \ref{Fig5} (b) it is seen that the bunching diminishes as the pump amplitude $F$ is increased which is due to the increased population of coherent single-particle states. Indeed, it is well-known that as the drive term becomes dominant in the Hamiltonian the system tends to a coherent state, well captured by the Gross-Pitaevskii approach. Note however that the dramatic bunching can remain even for $F$ much larger than $U$ (for example for the parameters considered in Fig. \ref{Fig5} we find $g_b^{(2)}\sim 10$ for $F/U \sim 10$)  and for $F$ larger than $\gamma$ (for example in Fig. \ref{Fig5} (b) we find $g_b^{(2)}\sim 2$ for $F/\gamma \sim 3$, with a photon density on the dark site $n_b \sim 0.1$). Fig. \ref{Fig5} (c) reveals slight anti-bunching for small $J$ while for larger hopping parameter $J$ an increasingly pronounced bunching is observed which eventually saturates in the limit $J \rightarrow +\infty$. This saturation can be understood by realising that a coherent population of the other resonances (at $\Delta=\pm\sqrt{2}J$, see Fig. \ref{Fig10} (b)) contributes to the dark site population. As $J$ is increased these other resonances are further detuned, thus resulting in a smaller coherent population on the dark site. As a function of the nonlinearity, as shown in Fig. \ref{Fig5} (d), the $g_b^{(2)}$ starts at $1$ for $U = 0$, then it sharply increases as a function of $U$, with a main peak around $U/\gamma \sim 0.2$. At larger $U$ other peaks are observed before it finally decreases to the hard core limit $g_b^{(2)} = 0$ for $U \rightarrow +\infty$. In Fig. \ref{Fig5} (a) we have also presented the third order correlation function $g_b^{(3)}$ which is also strongly peaked around $\Delta = 0$ and which shows that multi-photonic states with more photons can be probed by the higher order correlation functions. Note that our results for the correlation functions are much larger than the value $g_{th}^{(n)} = n!$ for thermal radiation \cite{Amann17072009}.

In the limit of weak pumping the steady-state of a driven-dissipative system becomes a pure state \cite{Carmichael199173,PhysRevA.83.021802,1367-2630-15-6-063001}. This state $\vert \Psi \rangle$ has a strong vacuum component $| 0 \rangle$, with corrections that can be calculated perturbatively with respect to the amplitude $F$, resulting in:
\begin{equation}
\vert \Psi \rangle =  \left| 0 \right> + \sum_j f_j^{(1)}\vert \Phi^{(1)}_j \rangle + \sum_j f_j^{(2)}\vert \Phi^{(2)}_j \rangle+ ...
\label{PureState}
\end{equation}
where  $\vert \Phi^{(n)}_j  \rangle$ denotes a basis  of the $n$-photon subspace and the amplitudes $f_j^{(n)} \sim \mathcal{O}(F^n)$. From the master equation the amplitudes $f_j^{(n)}$ can be calculated analytically for small $n$. The result for the second order correlation function of the state $\vert \Psi \rangle$ is presented as the dashed line in Fig. \ref{Fig5} (a). We have also verified the convergence of the numerical results to this value in the limit $F \to 0$. In order to better understand the role of the two-photon eigenstates given in Eqs. (\ref{EqES}) we consider the overlap of the state $\left| \Psi \right>$ with the six biphoton eigenstates $\left| \Psi^{(2)}_i \right>$ of the closed system in the limit $U/J \to 0$. This gives access to the probabilities $P^{(2)}_i$ of these two-photon eigenstates in the limit $F \to 0$ (normalized to the two-photon manifold), namely
$
P^{(2)}_i = \left|\langle \Psi^{(2)}_i| \Psi \rangle\right|^2/\sum_j \left|\langle \Psi^{(2)}_j| \Psi \rangle\right|^2 .
$
These probabilities are presented in Fig. \ref{Fig10} (d) for the two states of Eqs.  (\ref{EqES}) together with the sum over the probabilities for the other four eigenstates (the total sum is one). This clearly shows that around zero detuning the two-particle manifold is dominantly populated by the states considered previously in Eqs. (\ref{EqES}), moreover it confirms that the population of these states results in the super-bunching and the anomalous peak in the density at the dark site.  

\subsection{Extended Lieb lattices}

In order to study the fate of the strongly correlated states in larger lattices, we perform numerical simulations for 1D Lieb lattices made with 12 unit cells (i.e. 36 lattice sites) and 2D Lieb lattices with $4\times4$ unit cells (i.e. 48 lattice sites), both with periodic boundary conditions. The 2D lattice is also a candidate for the investigation of multi-photonic states since its spectrum also contains a flat band whose eigenstate displays dark sites \cite{PhysRevB.81.041410}. For these extended systems, the dimension of the Hilbert space becomes prohibitively large for an exact integration of the master equation due to the exponential growth with the number of sites. We have applied the corner-space renormalization method, which has been recently developed for driven-dissipative lattice systems \cite{PhysRevLett.115.080604}. In Table \ref{TableCorner}, we show the corresponding results. For the range of parameters studied, we notice that the $b$-sites have the same features as found for a single unit cell, i.e. small population and super-bunching effect. In particular, we notice that the local second order correlation function $g_b^{(2)}$ increases as the ratio $n_b/n_a$ between the populations of the dark and of the bright sites decreases. The strongest correlations are obtained for small pump intensity $F$ and large tunnelling $J$ between the sites. Comparing the 2D results with those for the 1D lattice we find similar qualitative effects, but with quantitatively weaker correlations.

\begin{table}
\begin{tabular}{ccc|cc|ccc}
 $N_{cells}$ & $N_{sites}$ & $M_{\mathcal{C}}$ & $F/\gamma$ &  $J/\gamma$ & $n_b/n_a$ & $g_b^{(2)}$ & $g_{<i,j>}^{(2)}$\\
\hline
\hline
\multicolumn{8}{c}{1D Lieb lattice}\\
\hline
12 & 36 & 3000 & 0.1 & 2   
   & 0.0180(5)   & 342(8)  & 19.3(4) \\
12 & 36 & 2000 & 0.1 & 1  
   &   0.0650(3)   & 23.3(2) & 2.35(2) \\

\hline
\multicolumn{8}{c}{2D Lieb lattice}\\
\hline
$4 \times 4$ & 48 & 5000 & 0.1 & 2 
             & 0.0161(1) & 66.2(2) & 1.42(3) \\
$4 \times 4$ & 48 & 5000 & 0.1 & 1 
             & 0.0631(1) & 4.41(1) & 0.996(2) \\
\hline
\end{tabular}

\caption{Results for the 1D and 2D Lieb lattices with periodic boundary conditions, obtained with the corner space renormalization method \cite{PhysRevLett.115.080604}. The value $M_{\mathcal{C}}$ indicates the dimension of the corner space needed to get the convergence of numerical results. The other parameters are $\Delta = 0$ and $U = 0.3 \gamma$. }
\label{TableCorner}
\end{table}

\begin{figure}[h!]
 \includegraphics[width=0.5\textwidth]{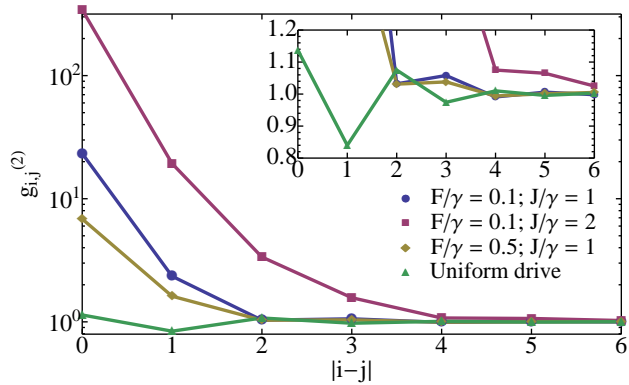}
  \caption{Second-order correlation function $g^{(2)}_{i,j}$ for dark sites  as a function of the distance $|i-j|$ between the sites (in term of unit cells) for a 1D Lieb lattice consisting of 12 unit cells and periodic boundary conditions. The same curves are shown in logarithmic scale (main figure) and in linear scale (inset). The green triangles are the results for a uniform drive with driving amplitude $F = 0.1 \gamma$ and hopping strength $J = 2\gamma$. The other parameters are the same for all results and are $U = 0.3 \gamma$ and $\Delta = 0$. \label{g2_ijCorner}}
\end{figure}

Importantly, the study of large lattices also allows us to investigate the correlations among photons on different dark sites, which can be described by the non-local second order correlation function $g_{i,j}^{(2)} =  \langle\hat{b}^\dagger_i\hat{b}^\dagger_j\hat{b}_j\hat{b}_i\rangle/\langle \hat{b}^\dagger_i\hat{b}_i\rangle^2$. We show in Table \ref{TableCorner} the results for the $g^{(2)}_{<i,j>}$ between two dark sites belonging to first-neighbor unit cells, and in Fig. \ref{g2_ijCorner}, we show the results for $g^{(2)}_{i,j}$ between two dark sites separated by $|i-j|$ unit cells. We find that, in general, the $g^{(2)}_{i,j}$ decreases to 1 as the distance between the sites increases with a superimposed oscillating behavior, typical for 1D driven-dissipative photonic systems \cite{PhysRevLett.104.113601, PhysRevLett.115.143601}. Nevertheless, we notice that in those regimes where the $b$-sites are particularly ''dark", there are significant correlations even for sites which are separated by 3 unit cells, i.e. 6 sites. This result indicates the possibility of probing long-range two-photon states even with a relatively weak nonlinearity. In Fig. \ref{g2_ijCorner}, we also display the result for $g^{(2)}_{i,j}$ obtained from pumping all the sites uniformly which clearly shows the disappearance of the dramatic bunching in this case.
 
\begin{figure}
  \includegraphics[width=0.5\textwidth]{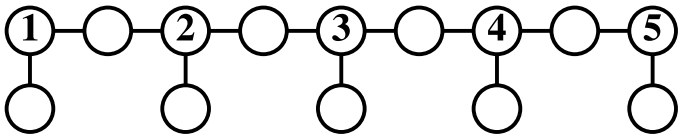}
  \caption{\label{Lieb1DOBC} Sketch of the 1D Lieb lattice with open boundary conditions studied here. The number $1, 2, \ldots, 5$ indicates the different dark sites.}
\end{figure}

Since experimental systems are more likely to be implemented with open boundary conditions, it is relevant also to see the effect of boundary conditions on the correlation effects discussed so far. For this reason, we have performed simulations for the 1D Lieb lattice made up of 14 sites sketched in Fig. \ref{Lieb1DOBC}, with partial driving and with open boundary conditions. In this case, because of boundary effects, the system is no longer homogeneous over the unit cells and the single particle eigenstates of the flat band obtain a small but non-zero energy and a finite population on the dark sites (i.e. the sites labeled $1,\ldots,5$ in Fig. \ref{Lieb1DOBC}). Nonetheless, we notice that these have the same features of the dark $b$-sites of the lattice with periodic boundary conditions: the mean photon density of these sites is much smaller than the density $\bar{n}$ of the brightest site in the lattice and the second order correlation function still show a strong bunching effect on these sites (see table \ref{TableOBC}). As in the case of periodic boundary conditions, we notice that even with open boundary conditions the $g^{(2)}_i$ is larger as the population $n_i$ of the dark site is lower.

\begin{table}
\begin{tabular}{c|ccccc}
$i$ & 1 & 2 & 3 & 4 & 5\\
\hline
$n_i/\bar{n}$ & 0.01514(5) & 0.00113(3) & 0.00565(3) & 0.00112(3) & 0.01515(5) \\
$g_i^{(2)}$ & 15.42(7) & 3940(130) & 111.4(8) & 4010(110) & 15.42(5) \\
$g_{i,1}^{(2)}$ & 15.42(7) & 5.11(7) & 1.45(2) & 2.78(3) & 1.014(1) \\
\hline
\end{tabular}
\caption{Results for the different dark sites of the Lieb lattice with open boundary conditions indicated in Fig \ref{Lieb1DOBC}. The non-local second order correlation function $g_{i,j}^{(2)}$ (third line of the table) is calculated with respect to the first dark site on the left of the lattice (site 1 in Fig. \ref{Lieb1DOBC}). The parameters of the simulations are: $F/\gamma = 0.1$, $U/\gamma = 0.3$, $J/\gamma = 2$, $\Delta = 0$.}\label{TableOBC}
\end{table}

The strong correlations for extended lattices can be interpreted in analogous way to what discussed in detail in Section \ref{Sec3Cav} for the three cavity case. The key difference is that the multiphoton states can involve photons belonging to distant unit cells. 
The anomalously large bunching on the dark sites is a consequence of the population of states with a finite probability of having two-photons on the same dark site. Since the coherent drive is on resonance with the flat band the coherent density is highly suppressed on these dark sites. The combination of these two effects results in the super bunching effect, observed in the local $g^{(2)}_{i}$ function in Table \ref{TableCorner} and Fig. \ref{g2_ijCorner}. The large values for the non-local $g^{(2)}_{i,j}$ between different dark sites are due to the population of states with a probability of having photons on both distant sites. 

\section{Conclusions}

In conclusion, we have presented a study of driven-dissipative cavity lattices with geometrical frustration and subject to partial driving, i.e. a fraction of the sites is undriven. 
One consequence of photonic lattices with geometric frustration is to produce single-particle states that have zero probability in dark sites. We have shown that in presence of relatively weak photon-photon interactions (interaction $U$ much smaller than the loss rate $\gamma$), the nominally dark states acquire an additional population of correlated photons, not captured by the Gross-Pitaevskii mean field approximation. We have presented a comprehensive study of 
lattices of the Lieb type. We have demonstrated analytically and numerically that, in spite of the moderate photon nolinearities, large correlations are visible due to multiphotonic resonances
involving the dark sites. 
The general paradigm presented in this letter is inherently of non-equilibrium nature: partial driving and frustration effects can lead to strongly correlated phases of light in cavity lattices, even when photon-photon interactions are relatively weak. These results should pave way to exciting investigations in a broad variety of photonic systems where photon-photon interactions are moderate. 

\section*{Acknowledgements}

We gratefully acknowledge discussions with A. Amo, F. Baboux, N. Bartolo, M. Biondi, J. Bloch. S. Finazzi and S. Schmidt. We  acknowledge  support  from  ERC  (via  the Consolidator  Grant ``CORPHO'' No.  616233),  from ANR (Project QUANDYDE No. ANR-11-BS10-0001), and from Institut Universitaire de France.

\bibliography{manusc}

\end{document}